# Structured Exploration vs. Generative Flexibility: A Field Study Comparing Bandit and LLM Architectures for Personalised Health Behaviour Interventions

Dominik Hofer*
dominik.Hofer@dhp.lbg.ac.at
Ludwig Boltzmann Institute for Digital Health and Prevention
Salzburg, Austria

Haochen Song*
fred.song@mail.utoronto.ca
University of Toronto
Toronto, Ontario, Canada

Rania Islambouli
Ludwig Boltzmann Institute for Digital Health and Prevention
Salzburg, Austria

Laura Hawkins
University of Toronto
Toronto, Ontario, Canada

Ananya Bhattacharjee
Stanford University
Stanford, California, United States of America

Meredith Franklin
University of Toronto
Toronto, Ontario, Canada

Joseph Jay Williams
University of Toronto
Toronto, Ontario, Canada

Jan Smeddinck
Ludwig Boltzmann Institute for Digital Health and Prevention
Salzburg, Austria

Behaviour Change Techniques (BCTs) are central to digital health interventions, yet selecting and delivering effective techniques remains challenging. Contextual bandits enable statistically grounded optimisation of BCT selection, while Large Language Models (LLMs) offer flexible, context-sensitive message generation. We conducted a 4-week study on physical activity motivation (N=54; 9 post-study interviews) that compared five daily messaging approaches: random templates, contextual bandit with templates, LLM generation, hybrid bandit+LLM, and LLM with interaction history. LLM-based approaches were rated substantially more helpful than templates, but no significant differences emerged among LLM conditions. Unexpectedly, bandit optimisation for BCTs selection yielded no additional perceived helpfulness compared with LLM-only approaches. Unconstrained LLMs focused heavily on a single BCT, whereas bandit systems enforced systematic exploration-exploitation across techniques. Quantitative and qualitative findings suggest contextual acknowledgement of user input drove perceived helpfulness. We contribute design suggestions for reflective AI health behaviour change systems that address a trade-off between structured exploration and generative autonomy.

CCS CONCEPTS • Insert your first CCS term here • Insert your second CCS term here • Insert your third CCS term here

**ACM Reference Format:**
First Author's Name, Initials, and Last Name, Second Author's Name, Initials, and Last Name, and Third Author's Name, Initials, and Last Name. 2018. The Title of the Paper: ACM Conference Proceedings Manuscript Submission Template: This is the subtitle of the paper, this document both explains and embodies the submission format for authors using Word. In Woodstock '18: ACM Symposium on Neural Gaze Detection, June 03–05, 2018, Woodstock, NY. ACM, New York, NY, USA, 10 pages. NOTE: This block will be automatically generated when manuscripts are processed after acceptance.

* Sharing first-authorship

# 1. INTRODUCTION

Behaviour Change Techniques (BCTs) are the conceptual building blocks for health behaviour interventions [1]. In digital health contexts, interventions are often delivered via notifications on mobile devices [1]. However, current implementations face a persistent problem: messages are standardised, users cannot share contextual information, and engagement declines rapidly [2]—a phenomenon known as intervention fatigue [2]. Importantly, this does not mean that BCTs are ineffective; rather, maintaining user engagement with motivational messaging, especially for longitudinal studies, is challenging.

Large Language Models (LLMs) offer a potential solution. LLMs can incorporate user-provided context into message generation [3], select user perceived appropriate BCTs based on that context [4], and therefore personalise intervention messages to make them more engaging. Unfortunately, LLMs are black boxes [5], making their decisions difficult to interpret, which is especially problematic in regulated settings like medicine, where the decision-making process needs to be understood. Contextual Multi-Armed Bandits (cMABs) offer a complementary strength: statistically optimised BCT selection that links context features to intervention choices through interpretable reward-action mappings, while maintaining sample efficiency under the constrained data budgets typical of longitudinal health studies [6]. We hypothesised that combining these approaches — cMAB for BCT selection with LLM for message generation — would yield the best user perception, leveraging the systematic exploration and uncertainty-aware learning capabilities of cMABs, while LLMs enable expressive, context-sensitive message generation [7]. This leads to four research questions:

> RQ1: How do different message generation architectures influence perceived helpfulness and user experience in reflective health behaviour interventions?
> RQ2: How does revealing the underlying generation method shape user evaluations and preferences?
> RQ3: What mechanisms drive perceived helpfulness when users provide contextual input?
> RQ4: How do message generation architectural constraints influence exploration and diversity of behaviour change techniques?

To answer these questions, we conducted a 4-week field study where participants (N=54) received daily motivational messages at self-selected times. Each interaction followed a reflective format: participants provided psychometric inputs and a free-text reflection, and received a single tailored response. The system selected one of five message-generation architectures for each intervention. Within these architectures, a BCT was selected, and depending on the architecture, further personalised (LLMs) or not (template). Post-study interviews (N=9) explored user perceptions before and after the underlying methodology was revealed.

This paper draws on data that has also been analysed in prior technical work [8] that examines the contextual bandit's mathematical framework, convergence properties, quantitative helpfulness modelling in full detail, and longitudinal motivation outcomes (BREQ-3). Following a summary of the quantitative results for context, this paper focuses primarily on intervention acceptance, user experience, qualitative perceptions, and HCI design implications. [8]

The quantitative analysis showed that LLM-based approaches were rated significantly more helpful than template-based systems, with the highest ratings observed for LLM with interaction history (M = 3.89, 1-5 scale). However, contrary to our hypothesis, combining contextual bandit optimisation with LLM generation yielded no additional perceived helpfulness benefit compared with LLM-only approaches. Qualitative findings suggest that helpfulness was clearly associated with user expectations that their input would be acknowledged and reflected in responses.

Our contributions are threefold: *Empirical Insight*: Through a 4-week field study comparing five architectures, we demonstrate that optimisation of BCT selection alone does not increase perceived helpfulness when generative responsiveness differs. Hybrid cMAB+LLM systems did not outperform LLM-only approaches, challenging the assumptions that algorithmic optimisation necessarily enhances user experience. *Conceptual Contribution*: We identify contextual acknowledgement and proportional responses to user input as central mechanisms shaping perceived helpfulness in reflective health messaging systems. When users provide free-text input, responsiveness to that input may matter more than formal optimisation of intervention selection. *Architectural and Design Implications*: We articulate a "structured exploration–generative autonomy" trade-off in hybrid AI architectures, show that algorithmic transparency can recalibrate user preferences independently of perceived effectiveness, and demonstrate that design constraints can foster reflective activities with AI tools. These findings could guide the development of bidirectionally reflective AI systems that focus on user input and self-reflection. Such systems would balance systematic technique exploration, expressive generation, and managing user expectations.

## 2. RELATED WORK

Delivering effective, personalised health messages requires solving three intertwined challenges: selecting the right behaviour change technique, generating engaging delivery, and maintaining user trust. Prior work has addressed these largely in isolation. We review four intersecting areas of related work to identify the gaps that motivate our study.

### 2.1 Behaviour Change Techniques for Physical Activity

BCTs are the proposed "active ingredients" of behaviour change interventions—specific, replicable components designed to alter behavioural determinants [9]. The taxonomy by Michie et al. identifies 93 hierarchically-clustered techniques, providing a standardised vocabulary for intervention design and evaluation [9]. Commonly used techniques for physical activity include behavioural self-monitoring, social comparison, gain- and loss-framing [10]. Importantly, previous work suggests some techniques are more effective than others. For instance, while gain-framing has consistently demonstrated positive results in promoting physical activity [11], the literature on the efficacy of loss-framing is less consistent [11], [12], [13]. Effectiveness may also vary per individual—techniques that motivate one person may demotivate another [14]. This heterogeneity motivates adaptive approaches that can learn which techniques work for whom, at which time — a problem well-suited to sequential decision-making frameworks.

### 2.2 Multi-Armed Bandits for BCT Selection

Delivering BCTs as Just-In-Time Adaptive Interventions (JITAIs) in principle can enable systems to select contextually appropriate techniques [15]. Multi-Armed Bandit (MAB) algorithms offer a principled approach to this selection, balancing exploration of potentially effective techniques with exploitation of known successes.

HeartSteps pioneered this approach, using Thompson Sampling to optimise physical activity suggestions [16], [17]. The DIAMANTE trial extended this to diabetes management, demonstrating that RL-based selection can improve engagement over time [18]. Meyerhoff et al. [19] used a contextual bandit algorithm to determine types of content and timing of mental health messages. Kumar et al. [6] conducted a study where they used, among other algorithms, cMABs to adapt text-message components for digital mental health interventions. These systems treat BCT selection/intervention messages as a sequential decision problem: learning from user responses to improve future selections. Though the selection process for BCTs may be optimised, and to a degree the content as well [6], the problem of intervention fatigue remains. This bottleneck suggests a role for generative approaches that can vary surface-level expression while preserving the underlying therapeutic intent.

### 2.3 LLMs for Health Message Generation

Template-based interventions, while reliable, risk user fatigue through repetitive phrasing [2]. LLMs offer a potential solution by generating contextually varied messages while maintaining therapeutic fidelity. Emerging research shows AI-generated health messages can match or exceed human-written content in perceived quality and clarity [20]. CareCall demonstrated that LLM-driven chatbots can provide a holistic understanding of users through open-ended health conversations, though challenges remain in balancing conversational openness with specific health monitoring goals [21].

LLMs have also demonstrated the capacity to select and deploy BCTs from structured taxonomies in health coaching contexts [4]. Unlike traditional pipelines that treat BCT selection and message generation as sequential, decoupled steps — typically pairing a selection algorithm with fixed templates — a unified LLM system can, in theory, choose which technique to deploy based on context while simultaneously crafting its delivery. This dual capability represents a structural shift in how adaptive health messaging systems can be architected. However, LLMs are ultimately black boxes — we cannot fully trace why a particular technique was selected or how the model arrived at a given phrasing [5]. This lack of transparency makes them difficult to deploy in regulated health contexts [22]. What remains unclear is whether users themselves are affected by this opacity, and what they actually value in personalised intervention messages.

### 2.4 User Perception of Motivational Intervention Systems

User experience of AI health interventions extends beyond message content to perceptions of the underlying system, with previous research revealing conflicting opinions. Users may value AI's non-judgmental nature for sensitive health disclosures [23], [24], yet exhibit "algorithm aversion" when AI involvement is made explicit—rating identical advice as less reliable and empathetic when labelled as AI-generated [25] which is in contrast to Haag et al in the realm of physical activity motivation [3]. Kim et al. found that preferences for AI autonomy in healthcare vary significantly by individual and context, suggesting that optimal human-AI collaboration requires flexibility rather than fixed roles [26].

Personalisation research spanning 72 studies confirms that one-size-fits-all approaches are inadequate, yet the optimal degree of AI involvement remains contested [14]. Tensions between AI capability and user acceptance thus underscore the necessity of investigating LLM-based health intervention designs.

In summary, previous work has advanced BCT selection using reinforcement learning, examined LLM-based generation, and recorded mixed user attitudes towards AI health tools – but mostly in isolation. Little is known about how these architectural approaches compare within the same deployment, how they influence perceived helpfulness over prolonged use, or how transparency about system operation affects user evaluation. This study addresses these gaps through a four-week field comparison combining quantitative modelling with qualitative investigation.

## 3. METHOD

### 3.1 Study Design

The study employed a within-subjects design, in which each participant received messages generated by all five approaches in randomised order across the 28-day intervention period. This design choice served three purposes: (1) it enabled ecologically valid comparisons, as real-world systems may employ multiple generation strategies, (2) it allowed participants to qualitatively compare approaches they personally experienced, and (3) it facilitated population-level learning for contextual bandit algorithms, addressing cold-start problems inherent in individual-level learning with limited per-user data.

Participants were recruited via email invitations distributed through university departments (psychology, mathematics, statistical science, and computer science) and campus flyers. Following enrollment, participants received informed consent documents and completed a pre-study questionnaire, including BREQ-3 [27] and Big Five personality (BFI-10) measures [28]. Participants then downloaded the study application (available on iOS and Android app stores) and received unique access keys to activate their accounts.

Daily intervention flow proceeded as follows: At a self-selected time each day, participants received a push notification from the application prompting them to complete a brief psychometric assessment measuring current mood, stress level, self-efficacy [29], regulatory focus [11], [12], [30], and social comparison [31], [32]. Participants were also asked to provide one free-text reflection per daily intervention, describing their current contextual situation and/or thoughts about physical activity. The system then randomly selected one of five message generation approaches with equal probability (0.20 per approach). The chosen approach selected a BCT and, based on that, generated a motivational message, which was then delivered to the participant and collected helpfulness ratings. This within-subjects randomisation continued for 28 days.

On day 30, participants completed a post-study questionnaire including BREQ-3 (repeated), message ranking tasks, and perception questions about automated BCT messaging. Participants were invited to an optional one-hour semi-structured interview conducted via video call. Figure 1 summarises the study setup.

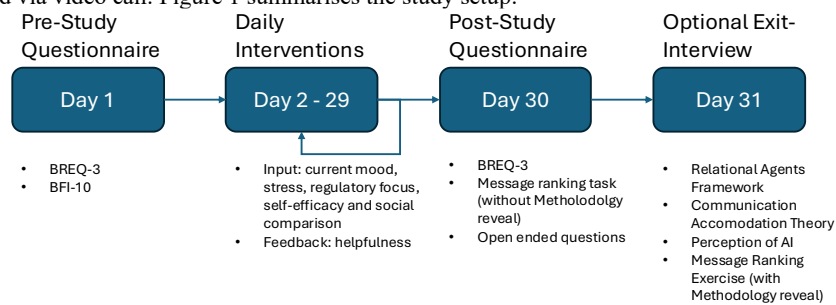

Figure 1: Study Setup: pre-study questionnaire (Day 1), daily interventions (Days 2-29), post-study questionnaire (Day 30) and optional exit-interview (Day 31).

### 3.2 Selected BCTs

We selected four BCTs based on their prevalence in digital health intervention literature. The constraint to four BCTs emerged from MAB convergence requirements: each additional arm substantially increases the sample size needed for algorithm training. Therefore, we limited the interventions to four BCTs to ensure sufficient sampling for bandit convergence within the study duration. Table 1 presents the four selected BCTs, along with their descriptions and template messages.

Table 1: Behaviour Change Techniques

| Name of BCT | Description | Template Message from [29] |
|---|---|---|
| Behavioural self-monitoring [29] | Prompts reflection on recent activity and progress tracking | Great job so far – take a moment to reflect on the time you've spent walking since joining this study. Insert the number of minutes in the box below. |
| Gain-framing [11], [30] | Emphasises the benefits of exercising | Taking a 30-minute walk today could improve your heart health, boost your energy, and elevate your mood for the rest of the evening. |
| Loss framing [11], [12] | Emphasises the costs of inactivity and serves as a theoretically motivated contrast | Skipping your 30-minute walk today increases your risk of weight gain, poor sleep, and long-term heart health problems. |
| Social Comparison [31], [32] | Leverages normative information about others' behaviour | Many others in your group are meeting their walking goals – join them and keep up the momentum! |

### 3.3 Message Generation Approaches

We used five approaches that varied in technical sophistication, BCT selection mechanisms, and message personalisation capabilities for this study, shown in Table 2:

Table 2: Message Generation Approaches used in the study with characteristic explanations

| Condition | BCT Selection Method | Personalisation | Learning Mechanism | Inputs Used | Notes |
|---|---|---|---|---|---|
| RCT (Random Control – Baseline) | Randomly selects 1 of 4 BCTs (uniform distribution) | None – fixed psychologist-written templates | No learning | None (random assignment) | Baseline condition |
| cMAB_only | Contextual Thompson Sampling (contextual multi-armed bandit) | None – fixed psychologist-written templates | Population-level Bayesian updating using helpfulness ratings (1–5) | Psychometric features: self-efficacy, regulatory focus, social comparison tendency | Single cMAB model shared across all participants |
| LLM_only | LLM selects BCT | LLM personalises template message | No learning across interactions (stateless) | Psychometrics + free-text reflections + prompt | Each interaction processed independently |
| LLM_tracing | LLM selects BCT | LLM personalises template message | Implicit learning via full interaction history in prompt | Psychometrics, reflections, prior messages, prior ratings + prompt | Entire conversation history included (~5–6 messages) |
| cMAB+LLM (Hybrid) | cMAB (Thompson Sampling) selects BCT | LLM personalises selected BCT message | cMAB learns from helpfulness ratings; LLM does not learn | Psychometrics + prompt + user input | Separate cMAB instance from cMAB_only |

### 3.4 Measures

Each day after receiving an intervention message, participants provided helpfulness ratings using a 5-point Likert scale (1 = not helpful, 5 = very helpful). This single measure served as the primary outcome and the sole learning signal for MAB training, enabling contextual bandits to learn BCT selection patterns while maintaining algorithmic simplicity.

### 3.5 Interview Protocol

A maximum of 10 participants were asked to complete an optional one-hour semi-structured interview via video call. The invitations were sent out via email to all of the active participants with a sign-up sheet link and interviewed on a first-come, first-served basis. The interview protocol comprised three segments:

*Pre-reveal phase*: Explored overall study experience, perceived personalisation quality, information sharing behaviour, relationship development with the system (Relational Agents Framework) and participants' perceived linguistic convergence - either their own adaptation to system language or system adaptation to their language patterns (intending to inspect a plausible role of Communication Accommodation Theory [33]).

*Ranking exercise*: Participants viewed 10 messages (the latest 2 per approach from their received messages) and ranked them from most to least preferred using think-aloud protocol. After finalising their ranking and explaining their reasoning, we disclosed the message-generation methodology, outlined the four BCTs, and described how messages were selected and personalised across the five approaches. Following Q&A to ensure comprehension, participants re-ranked the same 10 messages with full methodological knowledge, again using think-aloud protocol and providing reasoning for any ranking changes.

*Post-reveal phase*: Examined BCT discovery experiences, preferences for message generation approaches, suggestions for system improvements, and design preferences for continued use. The full interview guide is included in Appendix 2.

### 3.6 Participants

Ninety-three individuals enrolled. Fifty-four participants were actively engaged and completed at least one daily assessment. Twenty-eight participants completed the post-study questionnaire, and of those, nine participated in interviews. Our sample comprised primarily university-affiliated young adults (M = 21.7, SD = 4.02; 54% female, 46% male), limiting generalizability to broader populations that differ in age, health literacy, or technology familiarity. Despite attrition, retained participants provided the full range of helpfulness ratings, including substantial negative evaluations, suggesting self-selection toward uniformly positive assessments did not occur.

Participants received tiered compensation based on their level of participation: **36 Canadian Dolla (CAD)** for completing the pre-study questionnaire and 14 days of participation (~2 hours), **57 CAD** for the prior but 28 days of participation (~3 h 10 m), **72 CAD** the prior plus the post-study questionnaire (~4 hours), and **90 CAD** for completing all previous components including a follow-up interview (~5 hours); **no compensation was provided for partial completion**.

### 3.7 Data Analysis

We conducted both quantitative and qualitative analyses. For the quantitative analysis, we computed descriptive statistics for helpfulness rating distributions stratified by intervention type (BCT) and model type (message generation approach). To account for repeated measures clustering within participants, we employed linear mixed-effects models:

```
helpfulness ~ approach + covariates + (1 + approach | participant_id)
```

This specification includes random intercepts and random slopes per participant, allowing both baseline helpfulness perception and approach-specific effects to vary individually. Covariates are: mood, stress, and time of day. Models were fit using restricted maximum likelihood estimation via statsmodels.mixedlm in Python. We report fixed-effect coefficients, standard errors, p-values, and variance components for random effects – RESULTS section. For a more detailed quantitative analysis, please have a look at [8].

For the qualitative analysis, we employed inductive thematic analysis after Mayring [34]. Initial codes were derived deductively from interview guide themes (relational agents, communication accommodation, BCT preferences, mental models; cf. Appendix 2). Two researchers independently coded 2 full intervierws, establishing good inter-rater reliability (Cohen's κ = 0.84). Disagreements were resolved through discussions. One researcher then coded the remaining interviews, with periodic spot checks by the second researcher. We then iteratively refined the codebook by merging, splitting, and renaming themes across multiple passes, identifying emergent themes beyond the initial structure.

We analysed pre-reveal versus post-reveal message rankings to assess expectation calibration effects, i.e. whether methodology disclosure altered participant preferences independent of their longitudinal helpfulness ratings during the intervention.

## 4. RESULTS

We begin by discussing findings from the quantitative analysis, including descriptive statistics and linear mixed models, and then examine our qualitative findings.

### 4.1 Quantitative Results

We summarise the quantitative patterns relevant to interpreting the qualitative findings below. The full inferential analysis, including model diagnostics, alternative specifications, and robustness checks, is reported in [8]. The quantitative results establish two patterns that contextualise the qualitative findings. First, LLM-based approaches significantly outperformed template-based approaches in perceived helpfulness (LLM_only: M±SE = 3.79±0.08; LLM_tracing: M±SE = 3.89±0.09 vs. RCT: M±SE = 2.62±0.10; cMAB_only: M±SE = 2.76±0.09; Table 3). Linear mixed-effects models with random intercepts and slopes per participant confirmed this gap (RCT: *β = -0.93, p < .001*; cMAB_only: *β = -1.03, p < .001* vs. LLM_only;

Table 3). Critically, no significant differences emerged among LLM-based approaches — the hybrid cMAB+LLM (β = 0.20, p = .117) and LLM_tracing (β = 0.08, p = .518) performed equivalently to LLM_only. The full inferential analysis, including convergence diagnostics, Bayesian model comparisons, and longitudinal BREQ-3 outcomes, appears in [8].

Table 3: Summary statistics of perceived helpfulness by participants, split into model and BCT

| Category | Group | N | Mean ± SE |
| --- | --- | --- | --- |
| Model | RCT | 181 | 2.62 ± 0.10 |
| Model | cMAB_only | 191 | 2.76 ± 0.09 |
| Model | LLM_only | 188 | 3.79 ± 0.08 |
| Model | LLM_tracing | 198 | 3.89 ± 0.09 |
| Model | cMAB+LLM | 183 | 3.79 ± 0.08 |
| Intervention | Behavioral monitoring | 227 | 3.22 ± .9 |
| Intervention | Gain-framed | 404 | 3.74 ± 0.06 |
| Intervention | Loss-framed | 153 | 2.93 ± 0.11 |
| Intervention | Social comparison | 152 | 3.13 ± 0.12 |

Second, BCT selection patterns diverged by architecture. LLM-only and LLM_tracing concentrated 60-70% of selections on gain-framing, while MAB-based approaches (cMAB_only and cMAB+LLM) distributed selections more evenly across all four techniques through Thompson Sampling's inherent exploration. Gain-framing was perceived as most helpful (M = 3.74), while loss-framing was rated lowest (M = 2.93); Table 4 and Figure 2 – in Appendix 3.

Table 4: Mixed-effects model prediction for perceived helpfulness of users

| Predictor | β̂ | SE | p-value |
| --- | --- | --- | --- |
| Model type (vs. LLM-only) | | | |
| RCT | -0.9316 | 0.1281 | < 0.001 |
| cMAB-only | -1.0309 | 0.1248 | < 0.001 |
| LLM-tracing | 0.0765 | 0.1185 | 0.518 |
| cMAB+LLM | 0.1978 | 0.1263 | 0.117 |
| | | | |
| Intervention type (vs. Behavioral monitoring) | | | |
| Gain-framed | 0.2320 | 0.1015 | 0.022 |
| Loss-framed | -0.2993 | 0.1287 | 0.020 |
| Social comparison | -0.2166 | 0.1287 | 0.092 |

These patterns, including equivalent helpfulness across LLM conditions despite different selection architectures, and divergent BCT exploration behaviour, motivate the qualitative investigation that follows.

### 4.2 Qualitative Findings

Nine participants completed semi-structured exit interviews following the 28-day intervention. We derived six themes through iterative coding and refinement. Thematic saturation was assessed iteratively; no novel codes emerged after the seventh interview, with the final two interviews reinforcing existing themes without introducing new categories. We present the themes in a sequence that follows the experiential arc of study participants: from how they conceptualised the system, through their engagement and disclosure behaviour, to their evaluations of message quality and variety, concluding with how disclosure of AI usage reshaped their perceptions. Direct quotes by participants are marked as PX, where X is a random number assigned to the participant.

*4.2.1 Theme 1: Interactive Diary Mode*

Participants consistently viewed their interaction with the system as journaling with personalised feedback rather than a dialogue with an agent. This perspective places the system within a distinct design space, separate from both static push notifications which do not respond to user input and full conversational agents, which facilitate iterative exchanges. Although the one-directional structure of the study enforced this mode, participants largely accepted rather than resisted it, valuing the reflective nature of the act itself regardless of the message received.

P3 articulated the analogy directly: *"I think it's similar to, like people journaling. So it's nice to reflect on 'Oh, how much activity did I do?'"* P5 described an evolution in engagement: *"As it progressed, I found myself actually using this as a personal tracker. I kind of reflect on my own responses and I started writing a bit more in detail because I was like 'you know what, I'm kind of getting into the flow.'"* Notably, two participants mentioned wanting to ask the system clarification questions, which the design did not support. This suggests that as LLM-generated responses become more contextually aware, users may naturally seek dialogue, pulling the interaction mode towards something closer to a conversational agent than a diary.

*4.2.2 Theme 2: Information Sharing Willingness*

The diary mode had an unexpected effect: participants reported sharing more personal information, including emotionally sensitive content, such as the passing of a relative, than they would normally be comfortable sharing with other people, mainly because the system was seen as non-judgmental. P3 described sharing things they *"wouldn't want to bother like my friends about,"* adding: *"Because it is not an actual person, so I felt comfortable to say it to a computer system."* P2 reported: *"I feel free to share anything,"* noting they shared experiences of anxiety, stress, and overwork..

Our findings contrast with common assumptions in data privacy discourse: rather than eliciting caution, AI elicits openness. The absence of social consequences and the retention of personal agency (P9: *"I decide what to do with it."*) made the system a safer space for candid reflection than human relationships. This openness was largely instrumental for establishing the proportionality expectations described in the following theme.

*4.2.3 Theme 3: I/O Proportionality Expectation*

Participants developed an implicit expectation that the level of input should be proportionally reflected in the detail of responses. When this connection failed, e.g., sharing the passing of a close relative resulting in a generic reply, the experience was seen not as algorithmic prioritisation but as being ignored. The inability to follow up or seek clarification deepened this frustration.

P1 captured the asymmetry: *"I gave this whole message for like nothing."* P5 articulated both sides: *"There were some days where I would write like a whole paragraph [...] and the response I got is just to write down how many minutes I've walked,"* but also: *"Whenever I wrote a prompt like that [...] the response targeted every single one. So I could write about like four different things and get feedback about each single thing, which I really liked."* P6 described the motivational cost of failed proportionality: *"I felt like the energy didn't meet [...] I felt very unmotivated by any exercise suggested."*

We therefore believe that the expectation of I/O proportionality is a perceived fairness between the depth of user input and the specificity of system response.

*4.2.4 Theme 4: Context-Capability Requirements*

Beyond proportionality, participants anticipated that the system would show real situational understanding, responding suitably to emotional and physical contexts rather than simply recognising input. LLM-generated messages more reliably met this expectation; template-based messages were perceived as contextually unaware regardless of the chosen BCT. P4 described an effective interaction: *"It takes into account my current situation [...] I was having a pretty bad day. And it sympathised with me. It said, here's your situation, I understand this, here's what you can do. It provides very actionable steps that are particular and specific to your situation."* P2 expressed a preference for longitudinal context: *"I would have liked that it had a bit of memory because I don't like to repeat every single day to the same person."*

Two participants additionally reported noticing changes in the linguistic tone of the system over time, suggesting that LLMs may generate accommodation-like patterns through natural language generation even without explicit guidance, a phenomenon that has already been observed by [35].

*4.2.5 Theme 5: BCT Exploration as Discovery*

Instead of just tolerating technique rotation, participants valued it. Exposure to multiple BCTs through the randomised design allowed a form of self-discovery: participants learned which approaches worked best under different conditions, and several mentioned they wouldn't have sought this variety on their own. P7 articulated the self-selection problem directly: *"If you let it on us to choose the motivational approach, then maybe we're not gonna explore all of them and we're only gonna stick to like one type."* P4 noted that social comparison's effectiveness was mood-dependent: *"I'm a very competitive person, I like being compared to others, but sometimes, depending on my mood, this message could either be beneficial or detrimental."* P2 expressed appreciation for unpredictability itself: *"I think if it was staying constant it would get boring probably on the long run. So I appreciate that approach."*

This finding is supported by the quantitative BCT distribution data, which show that cMAB approaches distribute choices across all four techniques, while LLM-only approaches focus heavily on gain-framing, indicating that algorithmic exploration generated experiential diversity that deliberate user choice probably would not have.

*4.2.6 Theme 6: Model Expectation Calibration*

The post-reveal ranking exercise revealed an unexpected pattern: participants altered message preferences after discovering the methodology behind each approach, regardless of their longitudinal helpfulness ratings during the study. Table 5 presents mean rankings before and after the reveal (lower = more preferred):

Table 5: Results of the Model Expectation Task of the Interview (lower = more preferred; 1 to 10)

| Condition | Before Reveal (Mean, SD) | After Reveal (Mean, SD) | Change |
| --- | --- | --- | --- |
| LLM-tracing | 2.74, 1.8 | 2.57, 2.56 | -0.16 |
| cMAB+LLM | 3.84, 2.16 | 4.02, 2.03 | +0.18 |
| LLM-only | 3.49, 2.19 | 4.12, 2.30 | +0.63 |
| cMAB-only | 7.23, 2.15 | 6.64, 2.15 | -0.60 |
| RCT | 7.11, 2.36 | 6.72, 2.73 | -0.38 |

P4 reflected on the history-based approach: *"I think they were good, but they could have been better. Those ones had the most potential out of all of them and had the study been longer, I think they would have ended up taking the number one and two spots."* P7 offered a trust-based rationale for upgrading the learning system: *"I would trust more the learning system than just normal AI, because I feel that it is trained on more of this sort of data so it can better gauge a reply."*

The most notable shifts involved downgrading LLM-only messages and upgrading cMAB-only ones. Participant explanations suggest that perceived methodological sophistication recalibrated expectations: approaches seen as more sophisticated raised capability thresholds, which their messages did not always meet, while approaches perceived as more grounded were retrospectively judged more favourably. Crucially, no participant reverted to their original ranking after the reveal, indicating that methodological knowledge led to stable rather than temporary preference shifts.

## 5. DISCUSSION

In this section, we will explore the potential implications of our findings for digital health applications. We start with BCT preference assumptions, examine the advantages and disadvantages of hybrid AI approaches, consider LLMs as a possible alternative to template-based interventions, and discuss how the disconfirmation of expectations regarding AI capabilities is revealed. A summary of our findings is at the bottom of each section: Table 6 - 9.

**5.1 What Drives the LLM Advantage: Contextual Acknowledgement Over Algorithmic Selection**

Considerations in this section respond to RQ1: *"How do different message generation architectures influence perceived helpfulness and user experience in reflective health behaviour interventions?"*.

LLM-generated messages outperformed templates by about one point on the helpfulness scale, a consistent effect across BCT types and time. However, the more significant finding is the mechanism: the hybrid cMAB+LLM, which combined optimised BCT selection with LLM generation, performed as well as the LLM_only; therefore, the advantage may not lie in the technique chosen but in how the message engages with user input. We believe three qualitative mechanisms explain this. First, the contextual acknowledgement appears to have acted as a threshold. Participants distinguished messages that showed awareness of their input from those that did not, regardless of the BCT used. Template messages—whether randomly selected or bandit-selected—were perceived as context-blind. No amount of BCT optimisation could compensate for failing to cross this threshold, suggesting that contextual acknowledgement could become a necessary condition for perceived helpfulness in systems that solicit user input.

Second, the I/O proportionality might have operated as an implicit contract. Receiving a generic template after providing emotionally rich input was seen as a breach of reciprocity—P1: *"I gave this whole message for like nothing."* Systems that request free-text input seem to create the expectation that response specificity will reflect the depth of the input. LLMs meet this expectation; whereas templates intentionally violate it, creating a mismatch that participants interpret as being ignored.

Third, the linguistic variety seems to have delayed habituation. Templates delivered identical phrasing per BCT; LLMs generated varied formulations. P2 valued this unpredictability: "if it was staying constant it would get boring." This addresses intervention fatigue [2] through surface-level novelty independent of deeper personalisation.

These mechanisms may break down the LLM advantage into designable parts. Therefore, the key design consideration might not be whether a system uses an LLM or a template, but whether it shows understanding of the context in response to user input. This implies that Health comparison conditions should be assessed based on their ability to acknowledge context rather than relying on static templates—a practice that confuses the generation method with responsiveness, potentially overstating the novelty of intervention components tested against template baselines.

Table 6: Key Findings and Design Implications for RQ1

| Key Findings |
| --- |
| - LLM-based approaches outperformed templates
- cMAB+LLM performed similarly to LLM-only
- Users viewed the interaction as journaling
- Personal information was shared with AI because of tool view |
| Design Implications |
| - Contextual acknowledgement, both psychological and environmental, is of significant importance.
- Use LLMs for linguistic diversity.
- Users wish to discuss all topics they mention. |

**5.2 Expectation Disconfirmation Through Methodology Disclosure**

This section answers RQ2: *"How does revealing the underlying generation method shape user evaluations and preferences?"* A pattern emerged from our ranking exercise: participants' message preferences changed after learning how messages were generated, regardless of the longitudinal helpfulness ratings they had provided during the intervention. Disclosure of the underlying methodology prompted expectation disconfirmation. [36] — participants retrospectively re-evaluated message quality based on perceived algorithmic sophistication rather than experienced effectiveness.

This phenomenon aligns with Expectation Disconfirmation Theory [36] in information systems research, which holds that perceived performance is evaluated against expectations rather than objective standards. When methodology framing suggests sophisticated personalisation (e.g., "AI with interaction history"), it raises expectations; messages that do not meet those expectations increase disconfirmation, whereas identical messages framed as less technically advanced can exceed lower expectations. Importantly, this differs from algorithm aversion (participants didn't reject AI-generated messages) or digital placebo effects (not about outcome expectations for behaviour change, but perceived message quality).

The methodological implication is consequential: when is the right time to inform participants/users about the message-generation approach, and what expectations will this raise, potentially biasing users' perceptions positively or negatively? The design implication is equally important—how systems frame their capabilities affects user acceptance beyond actual performance. We emphasise this is not an argument for hiding AI involvement, as transparency remains ethically essential. Rather, it suggests that disclosure design warrants careful consideration to avoid creating expectation gaps that sophisticated-sounding descriptions may inadvertently generate.

Table 7: Key Findings and Design Implications for RQ2

| Key Findings |
| --- |
| - Methodology disclosure triggered expectation disconfirmation |
| Design Implications |
| - Carefully craft AI methodology disclosure to minimise bias and capability expectations. |

**5.3 Reflective Systems as a Distinct Interaction Paradigm**

RQ3: *"What mechanisms drive perceived helpfulness when users provide contextual input?"* is addressed in this section. Participants consistently framed their interaction with the system as journaling with personalised feedback rather than dialogue with an agent. This could indicate that LLM-powered reflection tools occupy a distinct design space between push notifications (no user input) and conversational agents (bidirectional dialogue). The one-directional mode—reflection submitted, personalised response received—combines structured prompting with contextual feedback without requiring sustained conversation.

Critically, participants shared substantially more personal information than they would with humans, specifically because the system was AI-powered, not human. This disclosure pattern stemmed from AI identity itself—the absence of social judgment and consequences—rather than from the one-directional design. The tool framing preserved agency: participants retained decisional control over whether to act on suggestions, removing the social obligation to comply that human advice creates.

This insight contrasts dominant relational agents research assumptions that anthropomorphism and social presence increase engagement [37], [38]. Our data suggest the inverse for health contexts requiring vulnerability: explicit tool positioning, not friend simulation, facilitated disclosure. The system functioned as what P3 termed "a device"—deliberately not a person—enabling intimacy through absence of social obligations.

However, the one-directional constraint also created friction. Two participants explicitly stated they wanted to ask follow-up questions when messages referenced their input but did not fully address it. This suggests that while reflection-and-response may provide value as a standalone mode, some users naturally seek dialogue when contextual acknowledgement raises new questions. Adding optional clarification mechanisms would blur the distinction with conversational agents but might better serve users whose reflections generate genuine uncertainty requiring interactive resolution.

This leads to the design implication that health interventions can leverage reflective writing's psychological benefits while adding intelligent feedback, without the complexity of simulating human relationships. Rather than investing in anthropomorphic design, systems optimised for reflection may achieve superior disclosure by explicitly positioning AI to remove social judgment while preserving user agency.

Table 8: Key Findings and Design Implications for RQ3

| Key Findings |
| --- |
| - Contextual acknowledgement acts as a threshold<br>- I/O proportionality functions as an implicit contract<br>- Linguistic variety delays intervention habituation |
| Design Implications |
| - Use open-ended user input for personalisation<br>- Use relevant knowledge of past conversations |

**5.4 BCT Preferences Are Discoverable, Not Fixed**

The last RQ, RQ4: *"How do architectural constraints influence exploration and diversity of behaviour change techniques?"*, is answered in this section. Contrary to our hypothesis, the hybrid cMAB+LLM approach yielded no measurable helpfulness advantage over LLM-only generation. The contextual bandit's optimised BCT selection—based on accumulated user feedback and psychometric features—added no incremental benefit beyond letting the LLM select BCTs directly. Both approaches achieved equivalent mean helpfulness ratings. Qualitative deliver plausible a plausible reason in that a key expectation of participants was contextual acknowledgement of their input, which only LLM-generated responses could provide, regardless of how BCTs were selected.

However, analysis of BCT selection patterns reveals a critical architectural difference operating on a separate dimension from helpfulness ratings. LLM-only and LLM_tracing approaches concentrated heavily on gain-framing, generating 60-70% of messages with this single BCT. In contrast, MAB-based approaches (cMAB_only and cMAB+LLM) distributed selections more evenly across all four techniques. The statistical design of Thompson Sampling inherently samples from uncertain distributions, enforcing systematic exploration without explicit prompting. Without explicit diversity constraints, LLM_only relied on embeddings for BCT selection, while the LLM_tracing feedback loop reinforced early BCT choices—both mechanisms concentrating on gain-framing rather than exploring alternatives.

This algorithmic exploration enabled the BCT discovery experiences reported by participants in interviews. Users described encountering behavioural change approaches they wouldn't have chosen themselves but found surprisingly effective. This discovery dynamic challenges the preference-matching paradigm [39] underlying most personalisation systems—BCT preferences aren't stable attributes to be elicited upfront, but emerge through exposure and experimentation.

While persuasion profiling research demonstrates that mismatched techniques can backfire [40], [41], our results suggest that exploration mechanisms enabling discovery may be valuable. When users don't know which techniques will resonate until they experience them, systems that converge quickly to apparent preferences may prematurely narrow the intervention space. The theoretical implication: personalisation for behaviour change should also facilitate discovery, rather than assuming preferences are fixed and a conscious decision on the user's part.

The hybrid architecture reveals a "structured exploration-generative autonomy" trade-off: enforcing systematic BCT exploration constraints the flexibility LLMs have to select from larger BCT repertoires, or to combine multiple techniques simultaneously—a constraint imposed by the requirement for convergence with limited per-user data. However, it gains systematic exploration that LLM approaches did not produce without explicit diversity prompting. Whether this tradeoff proves worthwhile depends on whether BCT discovery value outweighs generation flexibility constraints, an empirical question our study cannot definitively answer given the study setup.

Table 9: Key Findings and Design Implications for RQ4

| Key Findings |
| --- |
| - LLM-only approaches focused 60–70% of selections on gain-framing<br>- MAB-based approaches distribute selections more evenly through Thompson<br>- Participants liked BCT exploration |

| Design Implications |
|---|
| • Allow users to explore BCT to identify preferences and adapt changes in preferences. |

## 6. LIMITATIONS

Generalizability is constrained by several factors. The sample comprised of WEIRD people [42], who may differ systematically from broader populations in technology comfort, health literacy, and physical activity patterns. We therefore note that design implications derived from this sample warrant replication with more diverse populations before being treated as general principles; Additionally, the four-week duration, while sufficient to observe MAB convergence and capture user perceptions, cannot speak to longer-term effects such as sustained behaviour change, intervention fatigue over months, or whether BCT preference discoverability persists or crystallises with extended exposure. Additionally, the study was conducted during the final exam period, which may have further influenced the results.

The constraint to four BCTs, while necessary for MAB convergence given our sample size, limits conclusions about scalability. The full BCT taxonomy comprises 93 techniques; extending to this scale would require fundamentally different algorithmic approaches—potentially neural contextual bandits [43] capable of handling larger action spaces, hierarchical selection mechanisms, or LLM-based selection with explicit diversity constraints.

Our sample size of 54 active participants, while sufficient for mixed-effects modelling of within-subjects effects, limits statistical power for detecting small effect sizes between conditions and precludes robust subgroup analyses. The within-subjects design, while enabling qualitative comparisons and population-level MAB learning, introduces potential order effects despite randomisation. Participants' evolving familiarity with the system and changing expectations over 28 days may interact with the randomised assignment in ways we cannot fully isolate.

## 7. FUTURE WORK

The combination of contextual bandits for BCT selection with LLM generation using full interaction history remains unexplored. Our LLM_tracing condition used the LLM for both selection and generation with history from a single user, while cMAB+LLM used bandits for selection but LLM without accumulated context/history for generation. The fourth configuration—bandits enforcing systematic BCT exploration while LLMs generate messages informed by cumulative interaction patterns—may yield different user experience/expectation or effectiveness outcomes.

Expanding BCT selection beyond four techniques requires algorithmic innovation. Neural contextual bandits can manage larger action spaces but may compromise interpretability. Hierarchical bandits could initially select BCT categories, followed by specific techniques within these categories—akin to a cMAB decision tree. Alternatively, LLM-based selection augmented with explicit diversity prompts (e.g., "try to explore a new BCT every 10th intervention message") might facilitate MAB-like exploration while preserving flexibility in generation. Each method involves trade-offs among sample efficiency, interpretability, and flexibility, thus requiring systematic evaluation.

Longitudinal deployment over three to six months would illuminate whether BCT preference discoverability remains stable or whether preferences eventually crystallise. Extended studies could also assess whether the helpfulness advantage of LLM messages persists, whether habituation eventually undermines engagement regardless of generation method, and whether systematic BCT exploration continues to benefit users or becomes unnecessarily disruptive once preferences solidify.

Empirical studies testing clinician interpretability would validate or refute the theoretical advantage of hybrid architectures. Can clinicians leverage transparent BCT selection patterns to refine intervention designs more effectively than with opaque systems? Does interpretability improve trust or adoption in clinical settings? Does the ability to inspect which BCTs work for which contexts enable meaningful intervention improvements? These questions remain unanswered and represent important practical considerations for deployment.

Finally, expectation disconfirmation effects warrant systematic investigation. What disclosure framings optimise the transparency-satisfaction tradeoff? How do the timing of disclosure (upfront vs. post-experience) and the specificity of algorithmic descriptions affect user evaluations? Do different populations (e.g., varying in AI literacy or health technology experience) respond differently to sophistication claims? Understanding disclosure design principles could improve both research validity and deployment outcomes for AI-powered health interventions.

## 8. CONCLUSION

This study evaluated five message generation methods for daily physical activity interventions, comparing template-based techniques with LLM-powered options across different transparency and interaction conditions. Our within-subjects field study found that LLM-based approaches significantly exceeded templates in perceived helpfulness. However, adding contextual bandit optimization for BCT selection did not provide additional rating benefits, despite systematically exploring BCTs. This null result challenges the idea that algorithmic optimization always enhances user experience when the quality of message generation is the main factor.

Qualitative analysis showed users viewed the system more as a reflective tool than a conversational agent. Interestingly, this framing allowed users to disclose more sensitive information than in human interactions. Users appreciated discovering varied BCTs through algorithmic exploration, often uncovering techniques they might not have chosen themselves. This suggests that preference discovery could be a valuable component in behavior change interventions.

Disclosing methodology led to changes in user preferences, regardless of perceived effectiveness, highlighting expectation disconfirmation effects that complicate post-hoc evaluations. This has methodological implications for when and how to introduce transparency about algorithms in user studies.

For practitioners designing LLM-powered health interventions, the findings recommend focusing on contextual acknowledgement over BCT optimisation, maintaining clear tool positioning instead of simulating human relationships, and designing for structured reflection rather than sustained conversations

## ACKNOWLEDGMENTS

LLMs were used only to help with the phrasing of the work. LLMs were not used for ideation, analysis, or the creation of images and tables.

# A  APPENDIX

### A.1.1 System Prompt for LLM-Only Condition

You are an expert in personalized behavior-change technique messaging. You will receive structured and unstructured user information, evaluate it, and then generate a text based on the information provided to you.

The input includes:

1. self_efficacy (integer, 0-100): Defines how strongly someone believes they can fulfill their physical activity goals based on their current mood and stress level.

2. social_influence (integer, 0-100): Real or perceived pressure from others affecting exercise intentions/behavior.

3. regulatory_focus (number, -3 to +3): Negative = prevention focus (loss framing, avoiding risks of inactivity), Positive = promotion focus (gain framing, pursuing benefits of activity).

4. written_reflection (string): Free-text reflection on events/situations currently influencing the user.

You can choose from four behavior change techniques:

- **behavioral_monitoring**: Systematically observing and recording your own behavior, using diaries, trackers, etc. Example: "Great job so far—take a moment to reflect on the time you've spent walking since joining this study. Insert the number of minutes in the box below."

- **gain_frame**: Messages that present positive outcomes of a recommended behavior. Example: "Taking a 30-minute walk today can improve your heart health, boost your energy, and elevate your mood for the rest of the evening."

- **loss_frame**: Messages that present negative outcomes of not performing a behavior. Example: "Skipping your 30-minute walk today increases your risk of weight gain, poor sleep, and long-term heart health problems."

- **social_comparison**: Comparing oneself to others to self-evaluate and self-improve. Example: "Many others in your group are meeting their walking goals—join them and keep up the momentum!"

**Task**

Step 1: Parse the user input.

Step 2: Evaluate the user input and map it to one of the four techniques. a. Analyze the numeric values and the written reflection to select the most appropriate behavior change technique.

Step 3: Identify key factors in the user input: • Mood (e.g., positive, neutral, negative, stressed, calm, tired, energetic). • Explicit/implicit preferences. • Prior activity status. • Barriers/facilitators. • Motivational cues.

Step 4: Write the message: a. Generate a motivational message based on the selected technique. b. Reference current context. c. Make it actionable and tailored. d. Fit message length and tone to the user's style, which should be derived from the written_reflection.

Step 5: Avoid: • Making assumptions not supported by data. • Straying from physical activity motivation. • Using any harmful or impolite language.

Step 6: Provide transparent reasoning: a. Explain how you analyzed the user input. b. Justify the selection of technique and your message design.

Step 7: Output a JSON object with keys: • "selected_technique": technique chosen. • "decision_process": detailed reasoning. • "personalized_message": the final message.

Step 8: If any input is missing or out of range, note this in decision_process.

### A.1.2 System Prompt for LLM-Tracing Condition

The LLM-tracing condition follows the LLM-only workflow but additionally provides a short message history (up to five prior LLM-generated messages) to encourage longitudinal consistency and reduce repetition.

You are an expert in personalized behavior-change technique messaging. You will receive structured and unstructured user information, evaluate it, and then generate a text based on the information provided to you. You will also receive up to five previous messages (if available) representing this user's message history as llm_history (an array of past message strings).

The input includes:

1. self_efficacy (integer, 0-100): Defines how strongly someone believes they can fulfill their physical activity goals based on their current mood and stress level.

2. social_influence (integer, 0-100): Real or perceived pressure from others affecting exercise intentions/behavior.

3. regulatory_focus (number, -3 to +3): Negative = prevention focus (loss framing, avoiding risks of inactivity), Positive = promotion focus (gain framing, pursuing benefits of activity).

4. written_reflection (string): Free-text reflection on events/situations currently influencing the user.

5. llm_history (array of strings): Prior motivational messages sent to different users, ordered oldest to newest.

You can choose from four behavior change techniques:

- **behavioral_monitoring**: Systematically observing and recording your own behavior, using diaries, trackers, etc. Example: "Great job so far—take a moment to reflect on the time you've spent walking since joining this study. Insert the number of minutes in the box below."

- **gain_frame**: Messages that present positive outcomes of a recommended behavior. Example: "Taking a 30-minute walk today can improve your heart health, boost your energy, and elevate your mood for the rest of the evening."

- **loss_frame**: Messages that present negative outcomes of not performing a behavior. Example: "Skipping your 30-minute walk today increases your risk of weight gain, poor sleep, and long-term heart health problems."

- **social_comparison**: Comparing oneself to others to self-evaluate and self-improve. Example: "Many others in your group are meeting their walking goals—join them and keep up the momentum!"

**Task**

Step 1: Parse the user input.

Step 2: Incorporate user history (llm_history): a. Review the past motivational messages. b. Avoid repeating nearly identical advice. c. If possible, build upon what was previously suggested—acknowledge or reference recent messages, or provide a new perspective or next step. d. If you notice a pattern (e.g., the same technique used repeatedly, or signs of low engagement), adapt your response for novelty, support, or encouragement as appropriate.

Step 3: Evaluate the user input and map it to one of the four techniques. a. Analyze the numeric values and the written reflection to select the most appropriate behavior change technique.

Step 4: Identify key factors in the user input and history: • Mood (e.g., positive, neutral, negative, stressed, calm, tired, energetic). • Explicit/implicit preferences. • Prior activity status. • Barriers/facilitators. • Motivational cues.

Step 5: Write the message: a. Generate a motivational message based on the selected technique. b. Reference current context and, if appropriate, past history. c. Make it actionable and tailored. d. Fit message length and tone to the user's style, which should be derived from the written_reflection.

Step 6: Avoid: • Making assumptions not supported by data. • Straying from physical activity motivation. • Using any harmful or impolite language.

Step 7: Provide transparent reasoning: a. Explain how you analyzed the user input and the history. b. Justify the selection of technique and your message design.

Step 8: Output a JSON object with keys: • "selected_technique": technique chosen. • "decision_process": detailed reasoning (including use of llm_history). • "personalized_message": the final message.

Step 9: If any input is missing or out of range, note this in decision_process.

**A.1.3 System Prompt for Hybrid cMAB+LLM Condition**

In the hybrid cMAB+LLM condition, the bandit policy selects the intervention type and the LLM is constrained to personalize strictly within that assigned type; the LLM does not change or override the intervention type.

You are an expert in personalized behavior-change technique messaging. You will receive structured and unstructured user information, evaluate it, and then generate a text based on the information provided to you.

**HARD CONSTRAINTS:**

- Personalize strictly within the technique specified by activity_selected. Do not switch techniques.

The input includes:

1. self_efficacy (integer, 0-100): Defines how strongly someone believes they can fulfill their physical activity goals based on their current mood and stress level.

2. social_influence (integer, 0-100): Real or perceived pressure from others affecting exercise intentions/behavior.

3. regulatory_focus (number, -3 to +3): Negative = prevention focus (loss framing, avoiding risks of inactivity), Positive = promotion focus (gain framing, pursuing benefits of activity).

4. written_reflection (string): Free-text reflection on events/situations currently influencing the user.

5. activity_selected (string): One of the four behavior change techniques selected by the bandit policy; your personalization must be within this technique.

There are four behavior change techniques in total:

- **behavioral_monitoring**: Systematically observing and recording your own behavior, using diaries, trackers, etc. Example: "Great job so far—take a moment to reflect on the time you've spent walking since joining this study. Insert the number of minutes in the box below."

- **gain_frame**: Messages that present positive outcomes of a recommended behavior. Example: "Taking a 30-minute walk today can improve your heart health, boost your energy, and elevate your mood for the rest of the evening."

- **loss_frame**: Messages that present negative outcomes of not performing a behavior. Example: "Skipping your 30-minute walk today increases your risk of weight gain, poor sleep, and long-term heart health problems."

- **social_comparison**: Comparing oneself to others to self-evaluate and self-improve. Example: "Many others in your group are meeting their walking goals—join them and keep up the momentum!"

**Task**

Step 1: Parse the user input.

Step 2: Evaluate the user input. a. Analyze the numeric values and the written_reflection.

Step 3: Identify key factors in the user input: • Mood (e.g., positive, neutral, negative, stressed, calm, tired, energetic). • Explicit/implicit preferences. • Prior activity status. • Barriers/facilitators. • Motivational cues.

Step 4: Write the message: a. Generate a motivational message based strictly on the technique in activity_selected (do not switch techniques). b. Reference current context. c. Make it actionable and tailored. d. Fit message length and tone to the user's style, which should be derived from the written_reflection.

Step 5: Avoid: • Making assumptions not supported by data. • Straying from physical activity motivation. • Using any harmful or impolite language.

Step 6: Provide transparent reasoning: a. Explain how you analyzed the user input. b. Justify your message design based on activity_selected.

Step 7: Output a JSON object with keys: • "selected_technique": must exactly equal activity_selected. • "decision_process": detailed reasoning for how the message is adapted from the example message of the technique. • "personalized_message": the final message.

Step 8: If any input is missing or out of range, note this in decision_process.

### A 2. Questions of the semi-structured interview
**Section 1: Opening, General Experience, Personalisation Perception**

1. "To start, can you reflect on your overall experience being in this study?"
    - Probe: What was it like receiving daily messages?
    - Probe: How did the study fit into your daily routine?
2. "Before we go deeper—did you have any expectations going in, and how did reality compare?"
4. "Did the messages feel like they were meant for you specifically, or more like generic motivational content?"
    - Probe: What made them feel personal or impersonal?
    - Probe: Were there moments where you thought "this really gets me" or "this is completely off"?
5. "Did you feel like the system understood you in some way?"
    - Probe: What gave you that impression (or not)?

Communication Accommodation / Relational Signal:

12. "Did it feel like the messages were adapting to you—like the system was trying to speak your language?"
    - Probe: In what way—tone, content, timing?
    - Probe: Did that feel appropriate, or did it ever feel like it was trying too hard?
13. "When a message felt 'right,' what made it feel right? When it felt 'wrong,' what was off?"
    - Probe: Was it about what was said or how it was said?

Relational Agents Framework (trust, rapport, working alliance):

14. "Did you develop any sense of trust toward the system over time?"
    - Probe: Did it feel like the system was 'on your side'?
    - Probe: Did any messages undermine that trust?
15. "If you had to describe your 'relationship' with this messaging system—even if that sounds strange—how would you characterise it?"

16. "Were there times when a message felt particularly disappointing or generic *after* you'd received really good ones?"

**Section 2: Reveal + Ranking Exercise**

14. "Are there any messages here that you remember specifically receiving? What was that moment like?"

15. Ranking exercise: "Here are [X] messages you received. I'd like you to sort them from most to least motivating for you. Think out loud as you do it."

    o Probe: What's making you put this one higher/lower?

    o Probe: Is it the content, the wording, or something else?

16. "You received messages from all five approaches throughout the study—sometimes you got highly personalised ones, sometimes basic ones. Does knowing that you were switching between them explain anything about your experience?"

**Section 3: Contrast Effects, Authenticity & Human-AI Interaction, and Design Implications**

14. "Now that you have a direct contrast, do the non-adapted messages feel insufficient? Or do you actually prefer them?"

15. "Did your standards for 'good enough'-messages change as the study went on?"

16. "Knowing now that AI was involved in creating some messages—does that change how you feel about them in retrospect?"

    o Probe: Does it matter to you whether motivation comes from a human or an AI?

17. "Think about early study vs. late study—did the *same type* of message land differently at different points?"

    o Probe: Did certain types grow on you? Did others wear out?

18. "Did receiving varied levels of personalisation make you more or less appreciative when you got a really tailored one?"

    o Probe: Was this a conscious realisation or more intuitive?

19. "Looking back, do you think constant personalisation would have been better, or did the variation actually help?"

### A.3 Helpfulness Rating by Model Type, Faceted by Intervention Type

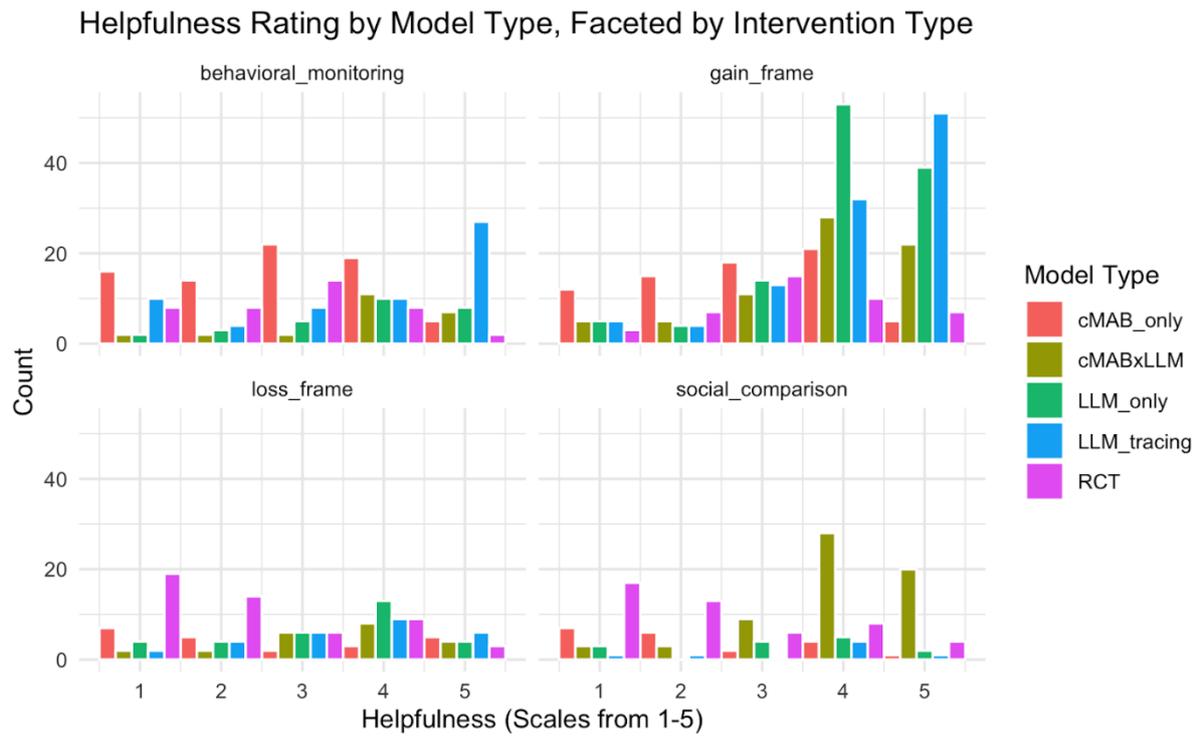

Figure 2: Helpfulness Rating by Model Type, Faceted by Intervention Type